# Formalizing Norm Extensions and Applications to Number Theory

## María Inés de Frutos-Fernández ✉ 🏠

Imperial College London, United Kingdom
Universidad Autónoma de Madrid, Spain

──── **Abstract** ────

The field $\mathbb{R}$ of real numbers is obtained from the rational numbers $\mathbb{Q}$ by taking the completion with respect to the usual absolute value. We then define the complex numbers $\mathbb{C}$ as an algebraic closure of $\mathbb{R}$. The $p$-adic analogue of the real numbers is the field $\mathbb{Q}_p$ of $p$-adic numbers, obtained by completing $\mathbb{Q}$ with respect to the $p$-adic norm. In this paper, we formalize in Lean 3 the definition of the $p$-adic analogue of the complex numbers, which is the field $\mathbb{C}_p$ of $p$-adic complex numbers, a field extension of $\mathbb{Q}_p$ which is both algebraically closed and complete with respect to the extension of the $p$-adic norm.

More generally, given a field $K$ complete with respect to a nonarchimedean real-valued norm, and an algebraic field extension $L/K$, we show that there is a unique norm on $L$ extending the given norm on $K$, with an explicit description.

Building on the definition of $\mathbb{C}_p$, we formalize the definition of the Fontaine period ring $B_{\mathrm{HT}}$ and discuss some applications to the theory of Galois representations and to $p$-adic Hodge theory.

The results formalized in this paper are a prerequisite to formalize Local Class Field Theory, which is a fundamental ingredient of the proof of Fermat's Last Theorem.

**2012 ACM Subject Classification** Theory of computation → Logic and verification; Theory of computation → Type theory

**Supplementary Material** The source code of the formalization is available at:
*Source code*: https://github.com/mariainesdff/norm_extensions_journal_submission

**Funding** EPSRC Grant EP/V048724/1: Digitising the Langlands Program (UK)

**Acknowledgements** I would like to thank Kevin Buzzard for many helpful conversations during the completion of this project, Thomas Browning for formalizing normal closures, and Yaël Dillies for the discussions on how best to integrate seminorms in `mathlib`. I also thank the `mathlib` community and maintainers for their support and insightful suggestions during the development of this work.

## 1 Introduction

Recall that the real numbers $\mathbb{R}$ are defined as the completion of the field $\mathbb{Q}$ of rational numbers with respect to the usual absolute value, and the complex numbers $\mathbb{C}$ as an algebraic closure of $\mathbb{R}$. The field $\mathbb{C}$ is algebraically closed and complete with respect to the extension of the usual absolute value.

However, there are other absolute values that we could consider on the rational numbers. Namely, for any prime number $p$, there is an associated $p$-adic absolute value on $\mathbb{Q}$ and, if we complete $\mathbb{Q}$ with respect to this absolute value, we obtain the field $\mathbb{Q}_p$ of $p$-adic numbers. Based on this definition, we can regard $\mathbb{Q}_p$ as an analogue of the real numbers $\mathbb{R}$.

The next natural question is which field would be the $p$-adic analogue of the complex numbers $\mathbb{C}$. Following the previous reasoning, a first candidate would be an algebraic closure $\mathbb{Q}_p^{\mathrm{alg}}$ of $\mathbb{Q}_p$. This field is algebraically closed, and we will see in Section 3 that it carries a well-defined extension of the $p$-adic absolute value. However, it turns out that $\mathbb{Q}_p^{\mathrm{alg}}$ is not complete with respect to this absolute value.

Therefore, to get an analogue of $\mathbb{C}$, we need the extra step of taking the completion of $\mathbb{Q}_p^{\mathrm{alg}}$ with respect to its absolute value, obtaining a new field $\mathbb{C}_p$, which is complete by construction and can be shown to be algebraically closed. We call $\mathbb{C}_p$ the field of $p$-adic complex numbers.

In general, given any field $K$ with an absolute value (or more generally, a seminorm, see Section 2) and a field extension $L/K$, we can ask whether the absolute value on $K$ can be extended to $L$ and, if so, whether this extension is unique. The main results in this article are the formalization of the proofs of two theorems, the Extension Theorem and the Unique Extension Theorem, giving positive answers to these questions, under hypotheses made precise in Section 3.

The construction of the field $\mathbb{C}_p$ of complex numbers is only one of the motivations for this formalization project. A second one is that this more general theory of norm extensions plays an important role in the proofs of the main theorems of local class field theory [2].

Class field theory is a branch of number theory whose goal is to describe the Galois abelian extensions of a local or global field $K$, as well as their corresponding Galois groups, in terms of the arithmetic of the field $K$. Class field theory is a key ingredient in the proof of the Taniyama–Shimura–Weil conjecture, which is in turn required to prove Fermat's Last Theorem.

A third motivation is that having formalized the field $\mathbb{C}_p$ opens the door to formalizing the definitions of *Fontaine's period rings* [19, 12], a work that we initiate in this paper (see Section 5.2). These rings have many applications to Representation Theory and $p$-adic Hodge Theory. They can be used to detect interesting properties of Galois representations, and to prove comparison theorems between different cohomology theories. They are studied within the Langlands Program [21], one of the biggest research programs in modern mathematics, consisting on a vast family of conjectures that seek to establish deep relations between algebra and analysis.

The results formalized in this paper are necessary prerequisites for the formalization of the proof of Fermat's Last Theorem, as well as for formalizing statements from the Langlands Program. They join a growing family of results from number theory and related areas formalized in Lean, including the $p$-adic numbers [22], Dedekind domains and class groups [6, 5], idèle class groups [15], Witt vectors [13], Galois theory [8], the Krull topology [25], the Haar measure [31], and perfectoid spaces [9]. See also Section 6.2 for a discussion of related work in other proof assistants.

A repository containing the formalization described in this article is publicly available at the link `https://github.com/mariainesdff/norm_extensions_journal_submission/`. In the article's text, `file.lean` represents the file at `https://github.com/mariainesdff/norm_extensions_journal_submission/blob/master/src/file.lean`. Parts of the formalization have already been integrated in Lean's mathematical library, in which case we will refer the reader to the corresponding files. Some of the code excerpts included in the paper have been edited for clarity.

## 1.1   Lean and mathlib

This formalization was carried out in the Lean 3 interactive theorem prover [16], based on dependent type theory, with proof irrelevance and non-cumulative universes [11]. An introduction to the language can be found in [3].

Our project is built on top of Lean's mathematical library `mathlib` [24], which currently contains over one million lines of code formalized by almost 300 contributors. The key property of this library is its unified approach to integrate different areas of mathematics, including for example algebra, analysis and topology, all of which we needed for this project. Both in Lean's core library and in `mathlib`, type classes are used to represent mathematical structures on types [4].

We remark that while Lean 4 is already available, the vast majority of the mathematical prerequisites for this project have not been ported to Lean 4's mathematical library yet, so at the current time it is not feasible to use this version for our work. We expect to port the project to Lean 4 once all of its prerequisites have been ported.

## 1.2 Paper outline

In Section 2 we recall some background on norms and field extensions. Section 3 contains an overview of the proofs of the main results of this paper, concerning the unique extension of a nonarchimedean norm. In Section 4 we discuss some implementation details of our formalization, while in Section 5 we present some applications of our main theorem to the field of Number Theory. Finally, we conclude in Section 6 with a discussion of future work and a reflection on the work presented in this article.

## 2 Mathematical background

In this section, we define several kinds of seminorms and norms on additive groups and rings, and we recall some definitions from field theory.

### Seminorms and norms

Let $G$ be an additive group. A *seminorm* on $G$ is a function $|\cdot| : G \to \mathbb{R}$ such that $|0| = 0$, $|-g| = |g|$ for all $g$ in $G$, and $|\cdot|$ is subadditive, that is, $|g + h| \leq |g| + |h|$ for all $g, h$ in $G$. A seminorm $|\cdot|$ such that $|g| = 0$ implies $g = 0$ is called a *norm*. An example of seminorm that is not a norm is the constant zero function.

```
structure add_group_seminorm (G : Type*) [add_group G] :=
(to_fun : G → ℝ)
(map_zero' : self.to_fun 0 = 0)
(add_le' : ∀ (g h : G), self.to_fun (g + h) ≤ self.to_fun g + self.to_fun h)
(neg' : ∀ (g : G), self.to_fun (-g) = self.to_fun g)

structure add_group_norm (G : Type*) [add_group G] extends add_group_seminorm G :=
(eq_zero_of_map_eq_zero' : ∀ g, to_fun g = 0 → g = 0)
```

We say that an additive group seminorm $|\cdot|$ is *nonarchimedean* if it satisfies the *strong triangle inequality*: $|g + h| \leq \max\{|g|, |h|\}$ for all $g, h$ in $G$. Note that this is stronger than the usual triangle inequality $|g + h| \leq |g| + |h|$. Otherwise, we say that $|\cdot|$ is *archimedean*.

```
def is_nonarchimedean {G : Type*} [add_group G] (f : G → ℝ) : Prop :=
∀ g h, f (g + h) ≤ max (f g) (f h)
```

If $R$ is a ring, then a *seminorm* on $R$ is an additive group seminorm $|\cdot|$ on $R$ that is also submultiplicative, that is, such that $|rs| \leq |r| \cdot |s|$ for all $r, s$ in $R$. A seminorm $|\cdot|$ is said to be *power multiplicative* if $|r^n| = |r|^n \ \forall r \in R, n \in \mathbb{N}_{\geq 1}$, and *multiplicative* if $|1| = 1$ and $|rs| = |r| \cdot |s|$ for all $r, s$ in $R$.

```
structure mul_ring_seminorm (R : Type*) [non_assoc_ring R] extends
add_group_seminorm R, monoid_with_zero_hom R ℝ
```

As in the additive group case, a ring seminorm $|\cdot|$ is a *norm* if $|r| = 0$ implies $r = 0$. A ring norm is said to be *power-multiplicative* or *multiplicative* if it has the corresponding property when regarded as a seminorm.

In this article, all rings will be assumed to be commutative and to have a unit element. We will only consider ring seminorms satisfying the extra hypothesis $|1| \leq 1$. This implies that either $|1| = 1$ or $|1| = 0$, in which case $|\cdot|$ is the zero seminorm.

An example of multiplicative ring norm on the rational numbers $\mathbb{Q}$ is given by the usual absolute value. This norm is archimedean, and if we complete $\mathbb{Q}$ with respect to it, we get the field $\mathbb{R}$ of real numbers.

However, there are other norms that we can consider on the rational numbers, which are widely used in Number Theory. Namely, for every prime number $p$, we can define a *p-adic norm* as follows. Define a function $v_p : \mathbb{Z} \to \mathbb{Z}$ as $v_p(r) := \max\{n \in \mathbb{Z} \mid p^n \text{ divides } r\}$. The function $v_p$ can be extended to $\mathbb{Q}$ by $v_p(\frac{r}{s}) = v_p(r) - v_p(s)$. We can then define the *p-adic norm* of $x \in \mathbb{Q}$ as $|x|_p := p^{-v_p(x)}$, and it is easy to check from its definition that $|\cdot|_p$ is a nonarchimedean multiplicative norm on $\mathbb{Q}$. When we complete $\mathbb{Q}$ with respect to the $p$-adic norm, we obtain the field $\mathbb{Q}_p$ of $p$-adic numbers.

Most of the definitions listed in this section have already been integrated in `mathlib` by the author, and can be found in the `mathlib` files `analysis/normed/group/seminorm.lean` and `analysis/normed/ring/seminorm.lean`.

We will be mainly interested in nonarchimedean (semi)norms, and in particular, the proof of the main theorem uses this property in a significant way. However, some intermediate results are true for arbitrary seminorms, and we have formalized them in that greater generality.

### Algebra norms

Let $R$ be a commutative ring with a (submultiplicative) norm $|\cdot|$ and let $A$ be an $R$-module. An *R-module seminorm* on $A$ is an additive group seminorm $\|\cdot\|$ on $A$ such that $\|r \cdot a\| = |r| \cdot \|a\|$ for all $r \in R, a \in A$. This notion was already defined in mathlib, under the name `seminorm`:

```
structure seminorm (R : Type*) (A : Type*) [semi_normed_ring R] [add_group A]
  [has_smul R A] extends add_group_seminorm A :=
(smul' : ∀ (r : R) (a : A), to_fun (r • a) = ‖r‖ * to_fun a)
```

If moreover $A$ is an $R$-algebra, we can define an *R-algebra norm* on $A$ as a ring norm $\|\cdot\|$ on $A$ such that $\|r \cdot a\| = |r| \cdot \|a\|$ for all $r \in R, a \in A$. This can be defined in Lean by extending the existing `seminorm` as follows:

```
structure algebra_norm (R : Type*) [semi_normed_comm_ring R] (A : Type*) [ring A]
  [algebra R A] extends seminorm R A, ring_norm A
```

### Field extensions

We end this section by recalling that given two fields $K$ and $L$, we say that $L$ is an *extension* of $K$, denoted $L/K$, if there is a homomorphism of rings $K \to L$ (which is necessarily injective). The extension $L/K$ is said to be *algebraic* if every element of $L$ is a root of a nonzero polynomial with coefficients in $K$. Given an extension $L/K$, $L$ is a vector space over the field $K$, and we say that the extension is *finite-dimensional* if the dimension of $L$ as a $K$-vector space is finite.

The following Lean code states that $L/K$ is a finite-dimensional algebraic extension of fields (the ring homomorphism from $K$ to $L$ is part of the data of the instance variable `[algebra K L]`).

```
variables {K L : Type*} [field K] [field L] [algebra K L]
  (h_fin : finite_dimensional K L) (h_alg : algebra.is_algebraic K L)
```

An algebraic field extension $L/K$ is *normal* if every irreducible polynomial in $K[X]$ that has a root in $L$ splits into linear factors as a polynomial in $L[X]$. Given any algebraic extension $L/K$, there is always a minimal field extension $N/L$ inside an algebraic closure of $L$ such that the extension $N/K$ is normal. The field $N$ is unique up to isomorphism, and we call it the *normal closure* of the extension $L/K$. If the extension $L/K$ is finite, then $N/K$ is finite as well.

## 3 Extensions of nonarchimedean norms

Let $K$ be a field with a nonarchimedean submultiplicative norm $|\cdot|$, and let $L/K$ be an algebraic extension. We would like to know whether it is possible to extend the norm $|\cdot|$ to a norm $|\cdot|_L$ on the larger field $L$ and, if so, whether this extension is unique. We will show in this section that both questions have a positive answer, under some conditions on the fields $K$ and $L$ and the starting norm $|\cdot|$. Moreover, we will provide an explicit description of the norm $|\cdot|_L$, which we will call the *spectral norm* induced by $K$. The main results we formalized are [7, Theorem 3.2.1/2 and Theorem 3.2.4/2].

### 3.1 The spectral norm

Let $R$ be a ring with a nonarchimedean seminorm $|\cdot|$, and let $P := X^m + a_{m-1}X^{m-1} + \cdots + a_1 X + a_0 \in R[X]$ be a monic polynomial of degree $m \geq 1$ with coefficients in $R$. The *spectral value* $\sigma(P)$ of $P$ is defined as $\sigma(P) := \max_{0 \leq i < m} |a_i|^{1/(m-i)}$. By convention, we say that the monic polynomial $P = 1$ of degree 0 has spectral value 0.

A first approach to formalize this definition would be to use the function `supr` to take the supremum of the values $|a_i|^{1/(m-i)}$ for $i$ running over the terms of `fin P.nat_degree`, the subtype of natural numbers less than the degree of $P$. However, this would force us to treat the case $P = 1$ differently in some of the proofs, since the type `fin P.nat_degree` is empty in that case.

Instead, we use the following trick: given any polynomial $P \in R[X]$, we define a map `spectral_value_terms` : $\mathbb{N} \to \mathbb{R}$ sending $i \in \mathbb{N}$ to $|a_i|^{1/(m-i)}$ if $i$ is less than the degree of $P$, or to 0 otherwise. Since every term $|a_i|^{1/(m-i)}$ is at least zero, taking the supremum of `spectral_value_terms` over all natural numbers returns the spectral value of $P$. Note that we do not ask that $P$ is monic in our formalized definition, but if $P$ is monic, both definitions agree.

```
variables {R : Type*} [semi_normed_ring R]
def spectral_value_terms (P : R[X]) : ℕ → ℝ := λ (n : ℕ),
  if n < P.nat_degree then ‖ P.coeff n ‖^(1/(P.nat_degree - n : ℝ)) else 0
def spectral_value (P : R[X]) : ℝ := supr (spectral_value_terms P)
```

We prove some of the basic properties of the spectral value, including:

1. The spectral value of a polynomial is always nonnegative.
2. The spectral value of the linear polynomial $X - r$ is equal to the seminorm $|r|$ of $r$.
3. For any $m \in \mathbb{N}$, the spectral value of $X^m$ is equal to 0. Moreover, if the seminorm $|\cdot|$ is a norm, then these are the only polynomials having spectral value 0.

Now, let $K$ be a field with a nonarchimedean submultiplicative norm $|\cdot|$, and let $L/K$ be an algebraic extension. Then any $y \in L$ is a root of a monic polynomial with coefficients in $K$, and the *minimal polynomial* of $y$ over $K$ is the monic polynomial of lowest degree in $K[X]$ having $y$ as a root.

The *spectral norm* $|\cdot|_{\mathrm{sp}}$ on $L$ is the function $|\cdot|_{\mathrm{sp}} : L \to \mathbb{R}_{\geq 0}$ sending $y \in L$ to the spectral value of the minimal polynomial of $y$ over $K$, which we will denote by $|y|_{\mathrm{sp}}$.

```
variables {K : Type*} [normed_field K] {L : Type*} [field L] [algebra K L]
  (h_alg : algebra.is_algebraic K L)
def spectral_norm (y : L) : ℝ := spectral_value (minpoly K y)
```

The terminology 'spectral norm' is justified by the fact, shown in the next subsection, that $|\cdot|_{\mathrm{sp}}$ is an algebra norm on $L$. However, note that this is not at all obvious from the definition, and both proving that the spectral norm satisfies the triangle inequality and that it is a multiplicative function require some serious work.

## 3.2 Norm extension theorems

In this section, we formalize in Lean 3 the proofs of the main results of the paper: two theorems about existence and uniqueness of extensions of nonarchimedean norms to algebraic field extensions.

First, we have the Extension Theorem, which states that given any field $K$ with a power-multiplicative nonarchimedean norm $|\cdot|$ and any algebraic field extension $L/K$, the spectral norm on $L$ is a power-multiplicative nonarchimedean $K$-algebra norm on $L$ extending the norm on $K$. The theorem also gives us information about how the spectral norm relates to the $K$-algebra automorphisms of $L$, and to other extensions of the norm to $L$:

▶ **Theorem 1** (Extension Theorem, [7, 3.2.1/2]). *Let $K$ be a field with a nonarchimedean power-multiplicative norm $|\cdot|$, $L/K$ an algebraic extension, and $G(L/K)$ the group of $K$-algebra automorphisms of $L$.*

- *The spectral norm $|\cdot|_{sp}$ on $L$ is a nonarchimedean power-multiplicative $K$-algebra norm on $L$ extending the norm $|\cdot|$ on $K$. All $K$-algebra isomorphisms of $L$ are isometries with respect to the spectral norm $|\cdot|_{sp}$. Any nonarchimedean power-multiplicative $K$-algebra norm on $L$ is bounded above by $|\cdot|_{sp}$.*

- *If the field extension $L/K$ is finite and normal, then $|\cdot|_{sp}$ is the only nonarchimedean power-multiplicative $K$-algebra norm on $L$ extending $|\cdot|$ for which all $g \in G(L/K)$ are isometries. If $|\cdot|'$ is a nonarchimedean power-multiplicative $K$-algebra norm on $L$ extending $|\cdot|$, then $|x|_{sp} = \max_{g \in G(L/K)} |g(x)|'$ for all $x \in L$.*

If moreover $K$ is complete with respect to a nonarchimedean multiplicative ring norm $|\cdot|$, then the spectral norm on $L$ is the *unique* nonarchimedean multiplicative ring norm on $L$ extending $|\cdot|$. This is called the Unique Extension Theorem.

▶ **Theorem 2** (Unique Extension Theorem, [7, 3.2.4/2]). *Let $K$ be a field that is complete with respect to a nonarchimedean multiplicative norm $|\cdot|$ and let $L/K$ be an algebraic extension. Then the spectral norm on $L$ is the unique multiplicative nonarchimedean norm on $L$ extending the norm $|\cdot|$ on $K$.*

We will now provide an overview of the proof of these theorems, referencing where to find the full details both in the literature and in our formalization. The general proof strategy is to perform a series of 'smoothing steps' in which, starting from a given seminorm (or norm), we construct a new seminorm or norm having better properties.

The proof of Theorem 1 relies on the following lemma:

▶ **Lemma 3.** *Let $K$ be a field with a nonarchimedean power-multiplicative norm $|\cdot|$. Each finite extension $L/K$ has at least one nonarchimedean power-multiplicative $K$-algebra norm extending the norm $|\cdot|$.*

The statement of this lemma is formalized as follows, and its proof can be found in the file `normed_space.lean`.

```
lemma finite_extension_pow_mul_seminorm (hfd : finite_dimensional K L)
  (hna : is_nonarchimedean (norm : K → ℝ)) :
  ∃ (f : algebra_norm K L), is_pow_mul f ∧ function_extends (norm : K → ℝ) f ∧
  is_nonarchimedean f := ...
```

**Proof of Lemma 3.** Fix a basis $\{e_1 = 1, \ldots, e_n\}$ of $L$ as a $K$-vector space and define a function $\|\cdot\| : L \to \mathbb{R}$ by setting $\|\sum_{i=1}^{n} a_i e_i\| := \max_i |a_i|$. We can check that $\|\cdot\|$ is a nonarchimedean $K$-module norm on $L$ extending the norm on $K$, and that there exists a positive real number $c$ such that $\|xy\| \leq c\|x\|\|y\|$ for all $x, y \in L$.

In the file `normed_space.lean`, we let `basis.norm` be the norm associated to a basis of a finite-dimensional $K$-vector space as above, and prove that it has the desired properties; in particular, the existence of the bounding constant $c$.

```
def basis.norm {ι : Type*} [fintype ι] [nonempty ι] (B : basis ι K L) : L → ℝ :=
λ x, ‖B.equiv_fun x (classical.some (finite.exists_max (λ i : ι, ‖B.equiv_fun x i‖ )))‖
lemma basis.norm_is_bdd {ι : Type*} [fintype ι] [nonempty ι] {B : basis ι K L} {i : ι}
  (hBi : B i = (1 : L)) (hna : is_nonarchimedean (norm : K → ℝ)) :
  ∃ (c : ℝ) (hc : 0 < c), ∀ (x y : L), B.norm (x * y) ≤ c * B.norm x * B.norm y := ...
```

The first smoothing step is to use [7, Proposition 1.2.1/2] to conclude the existence of a $K$-algebra norm on $L$ extending the norm $|\cdot|$ on $K$. We prove this proposition in the file `seminorm_from_bounded.lean`. That is, given a function $f : R \to \mathbb{R}$ from a commutative ring $R$ to the real numbers, we define a function `seminorm_from_bounded'`: $R \to \mathbb{R}$ by sending $x \in R$ to $\sup\left\{\frac{f(x \cdot y)}{f(y)} | y \in R, f(y) \neq 0\right\}$, and then prove that this function is a nonarchimedean ring seminorm whenever $f$ satisfies the required hypotheses.

Note that we do not need to make the condition $f(y) \neq 0$ explicit in the definition of `seminorm_from_bounded'`, since by Lean's convention $\frac{f(x \cdot y)}{f(y)}$ will be zero in that case, and we are only interested in this function when $f$ is a function taking nonnegative values.

```
def seminorm_from_bounded' : R → ℝ := λ x, supr (λ (y : R), f(x*y)/f(y))
def seminorm_from_bounded (f_zero : f 0 = 0) (f_nonneg : ∀ (x : R), 0 ≤ f x)
  (f_mul : ∃ (c : ℝ) (hc : 0 < c), ∀ (x y : R), f (x * y) ≤ c * f x * f y)
  (f_add : ∀ a b, f (a + b) ≤ f a + f b) (f_neg : ∀ (x : R), f (−x) = f x) : ring_seminorm R :=
{ to_fun    := seminorm_from_bounded' f,
  map_zero' := seminorm_from_bounded_zero f_zero,
  add_le'   := seminorm_from_bounded_add f_nonneg f_mul f_add,
  mul_le'   := seminorm_from_bounded_mul f_nonneg f_mul,
  neg'      := seminorm_from_bounded_neg f_neg }
lemma seminorm_from_bounded_is_nonarchimedean (f_nonneg : ∀ (x : R), 0 ≤ f x)
  (f_mul : ∃ (c : ℝ) (hc : 0 < c), ∀ (x y : R), f (x * y) ≤ c * f x * f y)
  (hna : is_nonarchimedean f) : is_nonarchimedean (seminorm_from_bounded' f) :=
```

The proof concludes with a second smoothing step, following [7, Proposition 1.3.2/1], which allows us to construct a power multiplicative $K$-algebra norm on $L$ extending $|\cdot|$. This proposition is formalized in the file `smoothing_seminorm.lean`. Given a real-valued function $f : R \to \mathbb{R}$ from a commutative ring $R$, we define a function `smoothing_seminorm_def` by sending $x \in R$ to the infimum $\inf_{n \geq 1} |x^n|^{1/n}$, which we show agrees with the limit of this sequence.

```
def smoothing_seminorm_def (x : R) : ℝ := infi (λ (n : pnat), (f(x^(n : ℕ)))^(1/(n : ℝ)))
lemma smoothing_seminorm_def_is_limit (hf1 : f 1 ≤ 1) (x : R) :
  tendsto (smoothing_seminorm_seq f x) at_top (𝒩 (smoothing_seminorm_def f x)) :=
```

We then prove that, whenever $f$ is any ring seminorm on $R$, the corresponding function `smoothing_seminorm_def` is a power-multiplicative ring seminorm on $R$. We remark that this smoothing step uses the nonarchimedean nature of the norm $|\cdot|$ in a significant way: proving that `smoothing_seminorm_def` satisfies the strong triangle inequality requires a careful approximation argument relying on $f$ being nonarchimedean. We are not aware of any alternative arguments to show that `smoothing_def` satisfies even the usual triangle inequality.

◄

Having proven Lemma 3, we can present the proofs of the two main theorems, whose formalizations can be found in the files `spectral_norm.lean` and `spectral_norm_unique.lean`.

**Proof of the Extension Theorem.** We want to show that the function $|\cdot|_{\mathrm{sp}} : L \to \mathbb{R}$ is a power-multiplicative $K$-algebra norm on $L$ extending the norm on $K$. We first reduce to the case where the field extension $L/K$ is finite and normal. We can do this because, to check that $|xy|_{\mathrm{sp}} \leq |x|_{\mathrm{sp}} |y|_{\mathrm{sp}}$, we can work on the normal closure of $K(x, y)$, and similarly for the other properties in the definition of power-multiplicative algebra norm. This reduction step just requires us to check that, whenever $E$ is an intermediate field between $K$ and $L$ and $x$ is an element of $E$, the spectral norm of $x$ is the same whether we regard it as an element of the normal closure of $E$, or as an element of $L$:

```
lemma spectral_value.eq_normal (E : intermediate_field K L)
  (h_alg_L : algebra.is_algebraic K L) (x : E) :
  spectral_norm K (normal_closure K E (algebraic_closure E))
    (algebra_map E (normal_closure K E (algebraic_closure E)) x) =
  spectral_norm K L (algebra_map E L x) := ...
```

Since $L/K$ is finite, by Lemma 3 there exists a power-multiplicative $K$-algebra norm $\|\cdot\|$ on $L$ extending the norm $|\cdot|$ on $K$.

The next 'smoothing step' is to define a function $|\cdot|_G : L \to \mathbb{R}$ that sends $y \in L$ to $|y|_G := |y|_{G(L/K)} := \max_{g \in G(L/K)} \|g(y)\|$. We prove in the file `alg_norm_of_galois.lean` that $|\cdot|_G$ (denoted `alg_norm_of_galois`) is a power-multiplicative $K$-algebra norm on $L$ extending the norm on $K$, and that every automorphism $g \in G(L/K)$ is an isometry with respect to $|\cdot|_G$.

```
def alg_norm_of_galois (hna : is_nonarchimedean (norm : K → ℝ)) :
  algebra_norm K L :=
{ to_fun   := λ x, (supr (λ (σ : L ≃ₐ[K] L), alg_norm_of_auto h_fin hna σ x)),
  ... }
lemma alg_norm_of_galois_is_pow_mul (hna : is_nonarchimedean (norm : K → ℝ)) :
  is_pow_mul (alg_norm_of_galois h_fin hna) :=..
```

Since the extension $L/K$ is normal, the minimal polynomial $q_y$ of $y \in L$ is of the form $q_y = \prod (X - g(y))^{p^e}$, where the exponent $e$ is a positive natural number depending on $y$. We can therefore use [7, Proposition 3.1.2/1(2)] to conclude that $|y|_{sp} = |y|_G$.

```
lemma spectral_norm_eq_alg_norm_of_galois (h_alg : algebra.is_algebraic K L)
  (h_fin : finite_dimensional K L) (hn : normal K L)
  (hna : is_nonarchimedean (norm : K → ℝ)) :
  spectral_norm K L = alg_norm_of_galois h_fin hna := ...
```

Hence, we have shown that the spectral norm on $L$ is a power-multiplicative $K$-algebra norm on $L$ extending the norm on $K$, and that for any other such norm $\|\cdot\|$, we have $|y|_{sp} = \max_{g \in G(L/K)} \|g(y)\|$ for all $y \in L$.                                                       ◄

**Proof of the Unique Extension Theorem.** We first show that the spectral norm $|\cdot|_{sp}$ is the only power-multiplicative $K$-algebra norm on $L$ extending $|\cdot|$. Suppose that $\|\cdot\|$ is another such norm. By [7, Prop. 3.1.5/1] , it suffices to check $\|\cdot\|$ and $|\cdot|_{sp}$ are equivalent on each field extension of the form $K(y)$, for $y \in L$. This follows from the facts that $K$ is complete and $K(y)$ is finite dimensional over $K$, and hence any two $K$-algebra norms on $K(y)$ will be equivalent.

```
theorem spectral_norm_unique' [complete_space K] {f : algebra_norm K L}
  (hf_pm : is_pow_mul f) (hna : is_nonarchimedean (norm : K → ℝ)) :
  f = spectral_alg_norm h_alg hna := ...
```

We point out an implementation detail of the proof of `spectral_norm.unique'`. In order to apply two existing `mathlib` lemmas about linear maps in this proof, respectively called `linear_map.continuous_of_finite_dimensional`, and `continuous_linear_map.is_bounded_linear_map` , we need to consider two different normed space structures on $K(y)$. However, we should not have two different `[normed_space K K(y)]` instances, since this would cause inference problems. To avoid this issue, we work with two copies of $K(y)$, each with their own normed space structure. We do this by defining a copy of $K(y)$ as `E := id K(y)`.

```
set E : Type* := id K(y) with hEdef
```

We use the identity map in this definition so that Lean is not able to infer a normed space structure on $K(y)$ from that on `E`. This allows us to put a different normed space structure on each of the copies.

```
letI N1 : normed_space K E := ...,
letI N2 : normed_space K K(y) :=...
```

To conclude the proof, we need to check that the spectral norm $|\cdot|_{sp}$ on $L$ is multiplicative, which requires a last 'smoothing step'. By [7, Proposition 1.3.2/2], for any $y \in L$, there exists a power-multiplicative $K$-algebra norm $|\cdot|_y$ on $L$ such that $y$ is multiplicative for $|\cdot|_y$, meaning that $|xy|_y = |x|_y |y|_y$ for all $x \in L$. This seminorm is defined by sending $x \in L$ to the limit of $\frac{f(xy^n)}{(f(y))^n}$ as $n$ tends to infinity, and can be found at `seminorm_from_const.lean`

```
def seminorm_from_const_seq (x : L) : ℕ → ℝ := λ n, (f (x * y^n))/((f y)^n)
def seminorm_from_const (x : L) : ℝ := classical.some
  (real.tendsto_of_is_bounded_antitone  (seminorm_from_const_is_bounded c f x)
  (seminorm_from_const_seq_antitone hf1 hc hpm x))
```

Since we have just shown that the spectral norm is the unique power-multiplicative $K$-algebra norm on $L$ that extends $|\cdot|$, we can conclude that $|\cdot|_{\mathrm{sp}} = |\cdot|_y$. Therefore every $y$ is multiplicative for $|\cdot|_{\mathrm{sp}}$, that is, the spectral norm is multiplicative.

```
lemma spectral_norm_is_mul [complete_space K]
  (hna : is_nonarchimedean (norm : K → ℝ)) (x y : L) :
  spectral_alg_norm h_alg hna (x * y) =
    spectral_alg_norm h_alg hna x * spectral_alg_norm h_alg hna y := ...
```

◀

The main reference we followed in our formalization, [7], is a book on nonarchimedean analysis, in which all results are stated exclusively for nonarchimedean (semi)norms. However, we would like to remark that the second smoothing step in the proof of Lemma 3 is the only part in the proofs of Theorems 1 and 2 in which the nonarchimedean property is necessary. By contrast, all of the remaining smoothing steps remain true for possibly nonarchimedean seminorms (noting that the extension of the norm will only be nonarchimedean if the starting norm has this property), and have been formalized in that generality.

We conclude this section with a concrete example: the extension of the $p$-adic norm on $\mathbb{Q}_p$ to its algebraic closure $\mathbb{Q}_p^{alg}$.

```
variables (p : ℕ) [fact (nat.prime p)]
@[reducible] def Q_p_alg : Type* := algebraic_closure ℚ_[p]
lemma Q_p_alg.is_algebraic : algebra.is_algebraic ℚ_[p] (Q_p_alg p) :=
algebraic_closure.is_algebraic _
```

By Theorems 1 and 2, the spectral norm is the unique (nonarchimedean) norm on the field $\mathbb{Q}_p^{alg}$ extending the $p$-adic norm.

```
instance normed_field : normed_field (Q_p_alg p) :=
@spectral_norm_to_normed_field ℚ_[p] _ _ _ _ padic.complete_space
  (Q_p_alg.is_algebraic p) padic_norm_e.nonarchimedean
lemma Q_p_alg.is_nonarchimedean : is_nonarchimedean (norm : (Q_p_alg p) → ℝ) :=
spectral_norm_is_nonarchimedean (Q_p_alg.is_algebraic p)
  padic_norm_e.nonarchimedean
lemma Q_p_alg.norm_extends (x : ℚ_[p]) : ‖ (x : Q_p_alg p) ‖ = ‖ x ‖ :=
spectral_alg_norm_extends (Q_p_alg.is_algebraic p) _ padic_norm_e.nonarchimedean
```

## 4    Implementation of norms and valuations

### 4.1   Unbundling seminorms

In `mathlib`, there are several classes to represent algebraic objects with a preferred norm that makes the object into a metric space. For example, a `normed_ring` $R$ is a ring endowed with a submultiplicative norm, which is used to define a metric space structure on $R$:

```
class normed_ring (R : Type u) : Type u :=
(to_has_norm : has_norm R)
(to_ring : ring R)
(to_metric_space : metric_space R)
(dist_eq : ∀ (x y : R), has_dist.dist x y = ‖x - y‖)
(norm_mul : ∀ (a b : R), ‖a * b‖ ≤ ‖a‖ * ‖b‖)
```

Other related classes are `semi_normed_ring`, `normed_field`, `normed_add_group`, etc. These classes are very useful to prove analytic results, provided that one only needs to consider one fixed norm in the algebraic object (group, ring, etc) being studied.

However, there are situations in which one needs to consider several seminorms on the same object. For example, the proof strategy for Theorems 1 and 2 consisted on, starting from a given seminorm on the field $L$, constructing a few other seminorms on $L$ having increasingly better properties.

The existing classes are not well-suited for working with several seminorms on the same object. The problem is that they bundle together the algebraic and topological structures of the object. For example, the above definition 'normed_ring' includes a field 'to_ring' that encodes the ring structure on $R$. If we were to put two `normed_ring` instances on $R$, this would in particular yield two distinct `ring` instances on $R$, which is not what we want - we want to consider two different norms on $R$, without varying the ring structure.

One could work around this problem by making multiple copies of the ring, as we did in the proof of Theorem 2. In that particular case, we decided on this approach because it allowed us to reuse some existing topological lemmas in `mathlib`.

However, we consider that a better general approach for simultaneously working with several norms on a ring $R$, which we follow in the rest of the paper, is to use unbundled versions of seminorms and norms. That is, instead of using a `normed_ring` class that bundles together the ring structure on $R$, its norm, and the resulting metric space structure, we work over a ring $R$ and we define the `ring_norm` as a function from $R$ to the real numbers satisfying the required hypotheses.

```
structure ring_norm (R : Type u) [ring R] : Type u :=
(to_fun : R → ℝ)
(map_zero' : to_fun 0 = 0)
(add_le' : ∀ (r s : R), to_fun (r + s) ≤ to_fun r + to_fun s)
(neg' : ∀ (r : R), to_fun (-r) = to_fun r)
(mul_le' : ∀ (x y : R), to_fun (x * y) ≤ to_fun x * to_fun y)
(eq_zero_of_map_eq_zero' : ∀ (x : R), to_fun x = 0 → x = 0)
```

## 4.2    Relating norms and valuations

A *valuation* $v$ on a ring $R$ is a multiplicative map $v : R → \Gamma_0$ to a linearly ordered commutative monoid with zero $\Gamma_0$ that preserves zero and one and satisfies the strong triangle inequality $v(x + y) \leq \max v(x), v(y)$ for all $x, y \in R$.

```
structure valuation (R : Type u) (Γ₀ : Type v)
  [linear_ordered_comm_monoid_with_zero Γ₀] [ring R] : Type (max u v) :=
(to_fun : R → Γ₀)
(map_zero' : to_fun 0 = 0)
(map_one' : to_fun 1 = 1)
(map_mul' : ∀ (x y : R), to_fun (x * y) = to_fun x * to_fun y)
(map_add_le_max' : ∀ (x y : R), to_fun (x + y) ≤ max (to_fun x) (to_fun y))
```

We say that a valuation $v$ has *rank one* if it is nontrivial and there exists an injective morphism of linear ordered groups with zero $\Gamma_0 → \mathbb{R}_{\geq 0}$.

```
variables {R : Type*} [ring R] {Γ₀ : Type*} [linear_ordered_comm_group_with_zero Γ₀]
class is_rank_one (v : valuation R Γ₀) :=
(hom : Γ₀ →*₀ ℝ≥0)
(strict_mono : strict_mono hom)
```

```
(nontrivial : ∃ r : R, v r ≠ 0 ∧ v r ≠ 1)
```

It is easy to see from these definitions that nontrivial nonarchimedean norms correspond to rank one valuations and, in practice, these terms are often used interchangeably in the mathematical literature. However, the formalization of these two notions in the library `mathlib` does not provide a way to relate them.

In the file `normed_valued.lean`, we formalize a dictionary between nonarchimedean norms and rank one valuations on a field $L$. This is a powerful tool, since it allows us to obtain full access to all of the theorems about these notions available in `mathlib`. Note that there are plenty of formalized results about normed fields and normed spaces, developed by analysts, as well as a very complete theory of valuations, mainly formalized as part of the perfectoid space project [9]. Without a way to convert between norms and valuations, we would be forced to make a choice about which of these results were available to us.

We relate the two definitions as follows. First, given a normed field $K$ for which the norm is nonarchimedean, this norm is automatically a valuation on $K$ (note that we need to use `nnnorm`, the version of the norm taking values on the type $\mathbb{R}_{\geq 0}$ of nonnegative reals).

We then define a function `normed_field.to_valued` that endows $K$ with a valued field structure. To do this, we need to provide a proof that the uniform space structure on $K$ induced by this valuation agrees with the one induced by the normed field structure.

```
variables {K : Type*} [hK : normed_field K]
include hK
def valuation_from_norm (h : is_nonarchimedean (norm : K → ℝ)) : valuation K ℝ≥0 :=
{ to_fun    := nnnorm,
  ... }
def normed_field.to_valued (h : is_nonarchimedean (norm : K → ℝ)) : valued K ℝ≥0 :=
{ v := valuation_from_norm h,
  is_topological_valuation := ...,
  ..hK.to_uniform_space,
  ..non_unital_normed_ring.to_normed_add_comm_group }
```

Conversely, if we start with a field $L$ with a valuation $v$ and a proof `hv` that $v$ is of rank one, then we can show that the function $L \to \mathbb{R}$ sending $x$ to the image of $v(x)$ under the homomorphism `hv.hom` is a nonarchimedean norm on $L$, and we can endow $L$ with the corresponding normed field structure.

Note that the default constructor of the class `normed_field` does not asks us to provide a uniform space structure on $L$; instead, it defines this uniform space structure as the one induced by the norm. However, doing this would lead Lean to think that we have two different uniform space structures on $L$, since we already had the uniform space structure induced by the valuation. We therefore indicate explicitly that the uniform space structure we are considering is the one coming from the valuation, and once again prove that this agrees with the one induced by the norm.

```
variables {L : Type*} [hL : field L] {Γ₀ : Type*} [linear_ordered_comm_group_with_zero Γ₀]
  [val : valued L Γ₀] [hv : is_rank_one val.v]
include hL val hv
def norm_def : L → ℝ := λ x : L, hv.hom (valued.v x)
def valued_field.to_normed_field : normed_field L :=
{ norm            := norm_def,
  dist            := λ x y, norm_def (x − y),
  to_uniform_space := valued.to_uniform_space,
  uniformity_dist := sorry,
```

```
...,  }
```

## 5   Applications to number theory

### 5.1   The $p$-adic complex numbers

As we recalled in Section 2, the real numbers $\mathbb{R}$ are constructed as the completion of the rational numbers $\mathbb{Q}$ with respect to the usual absolute value. We can then define the complex numbers $\mathbb{C}$ as an algebraic closure of $\mathbb{R}$. The field $\mathbb{C}$ is algebraically closed and complete with respect to the extension of the usual absolute value.

If we take the completion of $\mathbb{Q}$ with respect to the $p$-adic norm associated to a prime number $p$, we obtain the field $\mathbb{Q}_p$ of $p$-adic numbers, which we can regard as an analogue of the real numbers $\mathbb{R}$.

We would also like to find a $p$-adic analogue of the complex numbers $\mathbb{C}$. Our first guess would be to consider an algebraic closure $\mathbb{Q}_p^{\mathrm{alg}}$ of $\mathbb{Q}_p$. However, although $\mathbb{Q}_p^{\mathrm{alg}}$ is algebraically closed and, as shown in Section 3, the $p$-adic norm extends uniquely to $\mathbb{Q}_p^{\mathrm{alg}}$, it turns out that $\mathbb{Q}_p^{\mathrm{alg}}$ is not complete with respect to the $p$-adic norm.

By completing $\mathbb{Q}_p^{\mathrm{alg}}$ with respect to this norm, we obtain a new field $\mathbb{C}_p$, which is by construction complete with respect to the $p$-adic norm, and can be shown to be algebraically closed. Hence $\mathbb{C}_p$ can be regarded as a $p$-adic analogue of the complex numbers.

To formalize the definition of $\mathbb{C}_p$, we start from the definition of $\mathbb{Q}_p^{\mathrm{alg}}$. At the end of Section 3.2, we saw that `Q_p_alg p` is a normed field, whose norm is the spectral norm extending the $p$-adic norm on $\mathbb{Q}_p$. We take advantage of our norm-valuation dictionary from Section 4.2 to show that `Q_p_alg p` is a valued field, and define `C_p p` as the uniform space completion of `Q_p_alg p`. We then introduce the notation $\mathbb{C}$_[p] for `C_p p`, which is consistent with the notation $\mathbb{Q}$_[p] used in `mathlib` for the $p$-adic numbers.

```
instance Q_p_alg.valued_field : valued (Q_p_alg p) ℝ≥0 :=
normed_field.to_valued (Q_p_alg.is_nonarchimedean p)
def C_p := uniform_space.completion (Q_p_alg p)
notation `ℂ_[`p`]` := C_p p
```

An alternative, mathematically equivalent approach would have been to define $\mathbb{C}$_[p] as the Cauchy completion of `Q_p_alg p`, which would not require to introduce the instance `Q_p_alg.valued_field`. However, then we would have had to prove that $\mathbb{C}$_[p] is a normed field whose norm extends that of `Q_p_alg p`.

On the other hand, by defining $\mathbb{C}$_[p] as the uniform space completion of the valued field `Q_p_alg p`, we get access to existing results in `mathlib` that allow us to immediately conclude that $\mathbb{C}$_[p] is a valued field, whose valuation extends the valuation on `Q_p_alg p`. Therefore, this is a concrete example in which our norm-valuation dictionary has allowed us to gain access to lemmas that would otherwise not have been available.

```
instance : field ℂ_[p] := uniform_space.completion.field
instance C_p.valued_field : valued (ℂ_[p]) ℝ≥0 := valued.valued_completion
instance : has_coe_t (Q_p_alg p) ℂ_[p] := uniform_space.completion.has_coe_t _
lemma C_p.valuation_extends (x : Q_p_alg p) : valued.v (x : ℂ_[p]) = valued.v x :=
valued.extension_extends _
```

Now that we have a `valued_field` instance on $\mathbb{C}$_[p], we just need to show that its valuation has rank one (which is easy, since for example the element $p$ has valuation $1/p$, different from 0 and 1) to gain access to the associated `normed_field` instance on $\mathbb{C}$_[p].

```
instance : is_rank_one (C_p.valued_field p).v := ...
instance : normed_field ℂ_[p] := valued_field.to_normed_field
```

Having the above results, it is now easy to conclude that the norm on `ℂ_[p]` extends the norm on `Q_p_alg p`, and that it is nonarchimedean. All of the results in this section can be found in the file `Cp_def.lean`.

```
lemma C_p.norm_extends (x : Q_p_alg p) : ‖ (x : ℂ_[p]) ‖ = ‖ x ‖ := ...
lemma C_p.is_nonarchimedean : is_nonarchimedean (norm : ℂ_[p] → ℝ) := ...
```

## 5.2    Fontaine's period rings

Let $K$ be a $p$-adic field (a finite extension of $\mathbb{Q}_p$) and let $G_K := \mathrm{Gal}(K^{\mathrm{alg}}/K)$ be the *absolute Galois group* of $K$, that is, the group of $K$-algebra automorphisms of an algebraic closure $K^{\mathrm{alg}}$ of $K$. A $p$-adic *Galois representation* is a continuous group homomorphism $\rho : G_K \to \mathrm{GL}(V)$, where $V$ is a finite dimensional $\mathbb{Q}_p$-vector space.

Galois representations are a fundamental object of study in number theory. A precise understanding of how they relate to other mathematical objects (such as elliptic curves and modular forms) was a key ingredient in the proof of Fermat's Last Theory, and remains an active area of research within the Langlands Program, an ambitious collection of conjectures that seek to establish deep relations between seemingly distant areas of mathematics.

Of special interest are those Galois representation that 'come from geometry', meaning that the vector space $V$ is a subquotient of the étale cohomology group of an algebraic variety. A famous conjecture by Fontaine and Mazur predicts sufficient conditions for when a Galois representation comes from geometry in this sense.

Fontaine's strategy was to construct period rings, which are rings that can detect interesting properties of Galois representations. More precisely, a *Fontaine period ring* is a topological $\mathbb{Q}_p$-algebra $B$ with a continuous linear action of $G_K$, with some compatible additional structures (such as a Frobenius map or a filtration), such that the subring $B^{G_K}$ is a field of points of $B$ invariant under the Galois action is a field, and such that the $B^{G_K}$-vector space $D_B(V) = (B \otimes_{\mathbb{Q}_p} V)^{G_K}$ is an interesting invariant of the Galois representation $V$. Given a period ring $B$, a Galois representation $V$ is called $B$-*admissible* if $\dim_{B^{G_K}} D_B(V) = \dim_{\mathbb{Q}_p} V$.

For different choices of $B$, being $B$-admissible is equivalent to the representation having a certain arithmetic property. We have formalized in `Fontaine_period_rings.lean` the definitions of the following period rings:

1. $B = K^{\mathrm{alg}}$. A Galois representation $V$ is $K^{\mathrm{alg}}$-admissible if the action of $G_K$ on $V$ factors through a finite quotient.

   ```
   def K_alg {K : Type*} [field K] [algebra ℚ_[p] K]
     (h_fin : finite_dimensional ℚ_[p] K) := algebraic_closure K
   ```

2. $B = \mathbb{C}_p$. A Galois representation $V$ is $\mathbb{C}_p$-admissible if the action of the inertia subgroup $I_K$ of $G_K$ on $V$ factors through a finite quotient. See Section 5.1 for the formalization.

3. $B = B_{\mathrm{HT}} := \mathbb{C}_p[X, X^{-1}]$. A Galois representation $V$ is said to be $B_{\mathrm{HT}}$-admissible, or *Hodge-Tate*, if the vector space $\mathbb{C}_p \otimes_{\mathbb{Q}_p} V$ can be decomposed as a product of the form $\mathbb{C}_p(\chi_{\mathrm{cycl}}^{n_1}) \oplus \cdots \oplus \mathbb{C}_p(\chi_{\mathrm{cycl}}^{n_d})$ for some $n_i \in \mathbb{Z}$, where $\chi_{\mathrm{cycl}}$ denotes the cyclotomic character.

   ```
   def B_HT := laurent_polynomial ℂ_[p]
   ```

Besides being used to detect interesting properties of Galois representations, Fontaine's period rings are also prominently used in comparison theorems between different cohomology theories.

## 6 Discussion

### 6.1 Future Work

The first future goals related to this project would consist on formalizing some well-known properties of the fields $\mathbb{Q}_p^{\mathrm{alg}}$ and $\mathbb{C}_p$. For example, we could show that $\mathbb{C}_p$ is algebraically closed and that $\mathbb{Q}_p^{\mathrm{alg}}$ is not complete with respect to its norm. Certain generalizations of these facts are proven in [7, Proposition 3.4.1/3] and [7, Lemma 3.4.3/1], respectively.

Similarly, it is possible to build on our formalization to prove Hensel's Lemma [7, Proposition 3.3.4/3] and Krasner's Lemma [7, Corollary 3.4.2/2], two fundamental results in $p$-adic analysis. We remark that there is an existing formalization of Hensel's lemma in `mathlib`, but only for the $p$-adic numbers; the version we propose would generalize it.

A slightly more ambitious goal is to formalize the definition of the Fontaine period ring $B_{dR}$. We propose the following strategy to formalize this definition. First, let $E$ be the pre-tilt of $\mathbb{C}_p$, that is, the limit $E := \varprojlim_{x \mapsto x^p} \mathcal{O}_{\mathbb{C}_p}/(p)$, where $\mathcal{O}_{\mathbb{C}_p}$ is the ring of integers of $\mathbb{C}_p$. Let $A_{\mathrm{inf}} := W(E)$ be the ring of Witt vectors of $E$, and let $B_{\mathrm{inf}}^+ := A_{\mathrm{inf}}[\frac{1}{p}]$ be the localization of $A_{\mathrm{inf}}$ away from $p$.

```
def E := pre_tilt ℂ_[p] (C_p.valued_field p).v 𝒪_ℂ_[p] (valuation.integer.integers _) p
def A_inf := witt_vector p (E p)
def B_inf_plus := localization.away (p : A_inf p)
```

The missing part of the formalization consists on constructing a canonical surjective ring homomorphism $\theta : B_{\mathrm{inf}}^+ \to \mathbb{C}_p$ (the `noncomputable!` tag is required to avoid a timeout, but we expect to be able to remove it when we provide the definition of `theta`):

```
noncomputable! def theta : ring_hom (B_inf_plus p) ℂ_[p] := sorry
lemma theta.surjective : function.surjective (theta p) := sorry
```

By general properties of Witt vectors, to construct this function $\theta : B_{\mathrm{inf}}^+ \to \mathbb{C}_p$, it is enough to define the 'sharp' map $\cdot^\# : E \to \mathcal{O}_{\mathbb{C}_p}$ sending $\xi := (\xi_0, \xi_1, \cdots) \in E$ to $\xi^\# := \lim_{n \to \infty} \hat{\xi}_n^{p^n}$, where each $\hat{\xi}_n$ is an arbitrary lifting of $\xi_n$ to $\mathcal{O}_{\mathbb{C}_p}$, and to study some properties of this map.

Once the definition of `theta` is formalized, we will be able to define the ring $B_{\mathrm{dR}}^+$ as the completion of $B_{\mathrm{inf}}^+$ with respect to the ideal $\ker(\theta)$, and the field $B_{\mathrm{dR}}$ as the field of fractions of $B_{\mathrm{dR}}^+$.

```
def B_dR_plus := uniform_space.completion (B_inf_plus p)
def B_dR := fraction_ring (B_dR_plus p)
```

Our final future goal is to formalize some basic properties of the Fontaine period rings $B_{HT}$ and $B_{dR}$, as well as some of their applications to representation theory.

### 6.2 Related Work

This project requires to combine results from several mathematical areas, including analysis, field theory, number theory, and topology, and we found Lean's mathematical library `mathlib` to be the most complete library in terms of the required prerequisites. However, some of the background results we needed are also available in other proof assistants.

The *p*-adic numbers were first formalized by Pelayo, Voevodsky, and Warren [27], in the Coq UniMath library. They were formalized in Isabelle/HOL in 2022 [14].

Coq's Mathematical Components library [23] contains a formalization of Galois theory, developed as part of the odd order theorem project [20]. The proof that every field admits an algebraically closed extension was first formalized in the proof assistant Isabelle/HOL [17]. Field theory constructions such as algebraic extensions [30], algebraic closures [29], and minimal polynomials [28] have recently been added to the Mizar Mathematical Library.

The Isabelle/HOL standard library includes the theory of real normed spaces, much of which has been translated to the complex setting in an Isabelle Archive of Formal Proofs entry by Caballero and Unruh [10]. However, to the author's knowledge, the generalization to normed spaces over arbitrary fields is still missing. Similarly, real and complex normed spaces have been formalized in Mizar (see for instance [26] and [18]). In Coq, the MathComp-Analysis library [1] extends the Mathematical Components library with topics in analysis, including in particular results about normed spaces.

## 6.3   Conclusion

We develop the theory of extensions of nonarchimedean norms to algebraic field extensions, and we build on this work to formalize the field $\mathbb{C}_p$ of *p*-adic complex numbers and some of Fontaine's period rings. This project fits in within the long-term goal of formalizing a complete proof of Fermat's Last Theorem in the general case. It is also a starting point for formalizing Galois representation theory and *p*-adic Hodge theory.

The formalization required about 5000 lines of code, of which about 1000 lines have been integrated in the `mathlib` library at the time of writing this article.